# Surface Lattice Resonances in 3D Chiral Metacrystals for Plasmonic Sensing

Mariachiara Manoccio|[1], Vittorianna Tasco|[1]*, Francesco Todisco[1], Adriana Passaseo[1], Massimo Cuscunà[1], Iolena Tarantini[2], Marco Esposito[1]*

1. CNR NANOTEC Institute of Nanotechnology, Via Monteroni, Lecce 73100, Italy
2. Department of Mathematics and Physics Ennio De Giorgi, University of Salento, Via Arnesano, Lecce 73100, Italy

E-mail: marco.esposito@nanotec.cnr.it; vittorianna.tasco@nanotec.cnr.it
| contributed equally to his work

Chiral lattice modes are hybrid states arising from chiral plasmonic particles assembled in ordered arrays with opportune periodicity. These resonances exhibit dependence on excitation handedness, and their observation in plasmonic lattices is strictly related to the chiroptical features of the fundamental plasmonic unit. Here, we show the emergence of chiral surface lattice resonances in properly engineered arrays of nanohelices, fully 3D chiral nano-objects fabricated by focused ion beam processing. By tuning the relative weight of plasmonic and photonic components in the hybrid mode, we analyze the physical mechanism of strong diffractive coupling leading to the emergence of the lattice modes, opening the way to the engineering of chiral plasmonic systems for sensing applications. In particular, we identify a coupling regime where the combination of a large intrinsic circular dichroism of the plasmonic resonance with a well-defined balance between the photonic quality factor and the plasmonic field enhancement maximizes the capability of the system to discriminate refractive index changes in the surrounding medium. Our results lay the foundation for exploiting circular dichroism in plasmonic lattices for high performance biosensing.

In the last years, Surface Lattice Resonances (SLR), hybrid states arising from collective excitation of plasmonic nanostructures in periodic lattices, emerged as a promising research field in nanophotonics[1,2]. Because of their strong and spectrally narrow optical response, SLRs have been studied as platforms for exploring light-matter interaction phenomena[3] or applied to color filters[4,5], light-emitting devices [6,7], lenses [8], and nonlinear devices [9–11].

A new paradigm for these platforms lays in the inclusion of chiroptical effects in the lattice, leading to the recent observation of chiral SLRs (c-SLRs)[12] and extrinsic chiral effects[13], with the potential to extend the optical properties gamut of hybrid lattice resonances. C-SLRs arise when the interaction between the resonant chiral localized plasmon (c-LP) in the unit cell and the lattice diffractive order (DO) overcomes the plasmon losses, entering the strong coupling regime. In this condition, the energy detuning between the individual modes defines the plasmonic/photonic contribution to the hybrid state. Such a concept holds the potential for empowering schemes of optical detection based on circular dichroism (CD), since one can simultaneously exploit the high quality factor of SLRs and the large near-field amplification (M) and low mode volume (V) of plasmonic nanoantennas[14–19].

Nanohelices (NH), as fully 3D nanostructures with chiral shape, exhibit a very large optical activity, even under normal incidence conditions[20]. Indeed, in our recent work [21] we described how ordered arrays of 3D NHs can build up a chiral metacrystal, with resulting chiroptical response ruled out by out-of-plane and in-plane lattice parameters. However, the generation of c-SLRs in such a metamaterial has not been fully investigated.

In the present work, we show the emergence of c-SLRs in a nanohelix-based metacrystal, and we study the optical properties of the resulting resonances as a function of the energy detuning between plasmonic and diffractive components. As a proof of concept, we discuss the applicability of such platform for sensing applications: combining experimental and theoretical analysis, we show how the system sensitivity is governed by the plasmonic/photonic fraction of the c-SLR, as a trade-off between the large quality-factor of the diffractive component and the large field enhancement of the plasmonic counterpart, reaching an overall sensitivity of 530 nm/RIU. Our results allow to envision the interplay between chirality and surface lattice resonances as a novel concept for integrated optical nanosensors with high performances.

The fundamental unit of our system is a single, right-handed (RH), Pt-based nanohelix with an external diameter of 300 nm and vertical pitch of 500 nm, realized by focused ion beam processing[22] on an ITO-coated glass substrate and replicated several times to form an ordered square array (figure 1a, details in caption). The chiroptical response of the single NH is related to a more efficient excitation of electric dipoles when the circularly polarized light (CPL) matches the structure handedness. The chiral dipoles give rise to collective interactions between neighboring helices, and to a selective extinction of CPL in the visible spectral range, with two opposite circular dichroism bands[21]. Figure 1b shows the extinction spectra of a 30x30 elements array with lattice period (LP) of 460 nm, under illumination with CPL at normal incidence, as collected in different refractive index ($n$) media, in particular air (n=1), DI water (n=1.3334) and oil (n=1.518). When measured in air, the extinction spectra show the fingerprints of a chiral LSPR, with a higher extinction for RCP light (matching the handedness of the helices), peaked at $\lambda_{RCP}$=528 nm. On the other hand, extinction is lower and less structured when interacting with the opposite handedness CPL, with a slightly blue-shifted peak. In both cases, the large spectral broadening can be inferred to the mixed material composition of the NHs [23].

When switching to water and oil, the optical impedance mismatch between the superstrate and substrate around the nanostructures is reduced, thus sustaining the radiative coupling among neighbor NHs and increasing the visibility of the diffractive orders, as shown in figure 1b[1,24]. At the same time, the increased dielectric constant of the environment induces a redshift of the plasmonic resonance[25] towards the energy of the DO excited at normal incidence, whose momentum-energy dispersion for a 2D squared lattice in an homogeneous medium is given by:

$$\lambda_{DO} = \frac{2\pi n}{\sqrt{\left(k_x+N_x\frac{2\pi}{LP}\right)^2+\left(k_y+N_y\frac{2\pi}{LP}\right)^2}} \qquad (1)$$

where $k_x$ and $k_y$ are the k-vector components of the incident light ($k_x$=$k_y$=0, for normal incidence), $N_x$ and $N_y$ represent the diffraction orders in the xy basis and $n$ is the environment refractive index [26]. As further discussed later and shown in Figure 1b, moving from air to water environment, the main RCP resonance redshifts, crosses the first order DOs at λ(1,0)=610nm and splits in two peaks, which can be attributed to the upper and lower polariton (UP and LP, respectively) branches of an SLR, with UP=500 nm and LP=660 nm. An analogue trend is observed for the opposite polarization handedness, but with less intense features, as a result of the chiral optical behavior of the single NH, establishing a circular polarization dependent excitation of the SLR, which we can define as a chiral SLR (c-SLR).

Within oil (Figure 3b, blue line), a further redshift of the main spectral features of the c-SLR is observed, together with the emergence of the higher order (±1,±1) DO at 510 nm. The spectral features are sharper and more visible, as compared to the water environment, because of the further reduced refractive index mismatch between superstrate and substrate.

Since it is known how the array size limits the possibility to observe SLRs[27,28], in figure S1, we also show how the c-SLR intensity is significantly reduced when switching to a smaller array of 20x20 elements, corresponding to a patterned area of 100 μm$^2$.

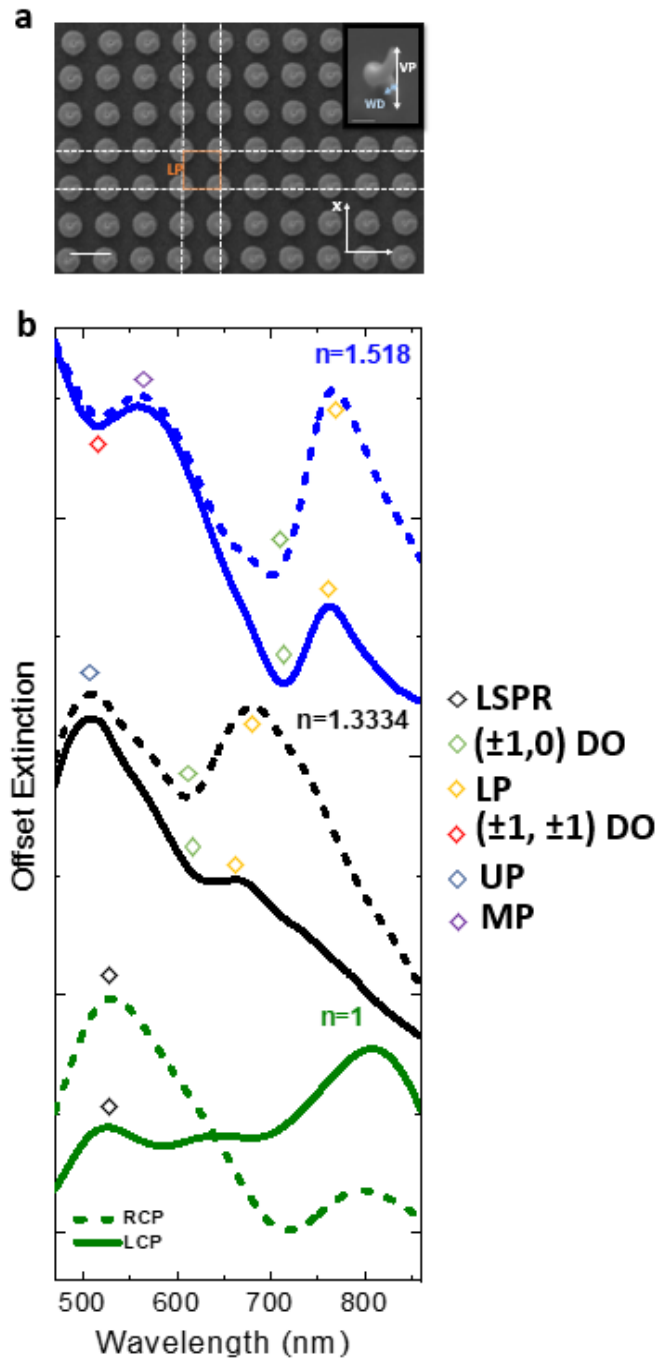

Figure 1. a. Top view SEM image of the nanohelices array with in-plane lattice period of 460nm and 30x30 elements. The scale bar is 500nm. The inset represents the SEM image of the single helix evidencing the vertical pitch (VP=500nm) and the wire diameter (WD=100nm). The scale bar is 200 nm. b. Left- (continuous line) and right-handed (dotted lines) circularly polarized extinction spectra of the array in air (n=1, green lines), water (n=1.3334, black lines) and oil (n=1.518, blue lines). The black symbols identify the helix localized plasmon. The diffractive orders are indicated with green (±1,0) and red (±1,±1) rhombs, respectively. The upper (UB), middle (MB) and lower (LB) branches of the circular surface lattice resonance are indicated by the navy, violet and yellow diamonds, respectively.

To shed light on the optical behavior of NHs lattices, we realized arrays with different LPs (400nm, 430nm, 460nm, 490nm). We experimentally measured the energy−momentum extinction maps by imaging the transmitted Fourier space under interaction with LCP and RCP light, in oil environment. The corresponding circular dichroism spectra, calculated as the difference between LCP and RCP extinction, are shown in Figure 2 (all the extinction spectra and Fourier space maps are reported for completeness in the supplementary information figure S2,3). Interestingly, although the DOs are linearly polarized, all the CD maps show the emergence of linear and parabolic dispersions directly related to the DOs excited in the lattice. In particular, for LP=400 nm, a clear band bending can be observed at the DOs (±1,0) due to the chiral LSPR of the NHs (figure 2 and S3). By increasing LP, the hybrid mode continues to be visible and becomes narrower. For the largest LP (490 nm), also the (±1,±1) DOs become more visible at higher energies and at normal incidence. As discussed above, RCP-SLRs appear more pronounced than LCP-SLRs (Figure S3), because of a more efficient excitation of the NH plasmon with RCP light[21]. This behaviour is confirmed in figure S4 by the simulated far-field extinction maps for LCP and RCP, in good agreement with the optical measurements.

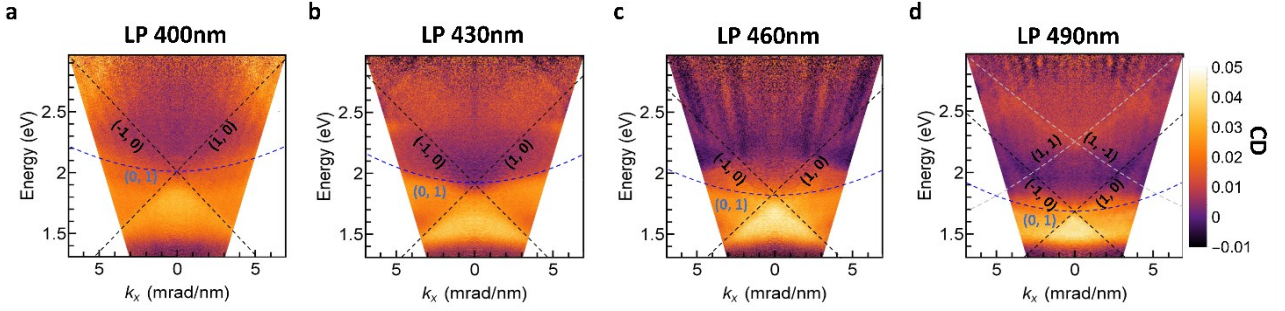

Figure 2. Energy-momentum experimental CD extinction maps defined as the difference between measured LCP and RCP extinction (shown in figure S3) at different LPs (400 nm, 430 nm, 460 nm, 490 nm). The dashed lines are drawn to highlight the progressive DO dispersion shift when increasing LP. The dispersion of the DOs were calculated from eq. 1 for the indicated (Nx,Ny) indices.

The experimentally measured spectral features (Figure 1) and the angular dispersion (figure 2 and S3) demonstrate the excitation of c-SLR in the visible range, arising from a properly engineered array of fully 3D chiral NHs. In particular, the proposed system exhibits a large value of maximum CD ($E_{LCP}-E_{RCP}$ larger than 5%) at the c-SLR energy (around 800 nm). To shed light on the optical behaviour of our system, we numerically calculated the normal incidence ($k_x = k_y = 0$) extinction of an infinite NHs array as a function of the lattice period and of the incident circular polarization in water environment, of significance for biosensing application. Such a study will show later how the detuning among plasmonic and DOs, controlled by the lattice period, affects the overall system sensitivity.

In order to estimate the bare LSPR position and to fully explore the diffractive coupling mechanism, we started the simulations from a very low LP value (300 nm) corresponding to NHs in closely packed configuration. Here, the reduced LP induces near-field coupling among adjacent NHs enhancing scattering and leading to stronger effective damping [29,30]. Therefore, the consequently larger plasmon mode results not suitable for sensing. Indeed, in the map, the RCP LSPR (Figure 3a) appears as a broad peak at around 2.7 eV (solid horizontal black line). The energy dispersions of the DOs are indicated with dashed black lines, as calculated from Equation 1.

By increasing the LP, as the first order DO energy crosses the LSPR (LP=350nm), an anti-crossing clearly appears in the LP-dependent map for both RCP (Figure 3a) and LCP (Figure S5d) light. The same behaviour is observed at LP=490 nm, when the LSPR crosses the second order (1,1) DO.

For each LP, we extracted the energy position of the extinction peaks by Gaussian profile fitting (represented as coloured dots in the maps). The obtained dispersion and anti-crossing behaviour, arising from the hybridization between the c-LSPR and the DOs, can be described by a three coupled oscillators model according to the following Hamiltonian:

$$\hat{H} = \begin{pmatrix} E_{DO_{(\pm1,0)}} & 0 & g^P_{(\pm1,0)} \\ 0 & E_{DO_{(\pm1,\pm1)}} & g^P_{(\pm1,\pm1)_2} \\ g^P_{(\pm1,0)} & g^P_{(\pm1,\pm1)} & E^P_{LSPR} - i\gamma \end{pmatrix} \quad (2)$$

Here, P represents the light polarization (LCP or RCP), the diagonal elements represent the diffractive wave energies of the (±1,0) and (±1, ±1) DOs, and the LSPR energy, while the off-diagonal elements, $g^P_{(\pm1,0)}$ and $g^P_{(\pm1,\pm1)}$, describe the coupling strength between the corresponding modes. $\gamma$ represents the plasmonic losses, while coupling between different DOs can be ignored[26]. The plasmon energy $E_{LSPR}$ and the parameters $g^P$ and $\gamma$ have been considered as fitting parameters.

By fitting the peaks position with the Hamiltonian solution, we found for RCP excitation, $E^{RCP}_{LSPR}=2,7$ eV, $g^{RCP}_{(\pm1,0)}\sim$ 340meV and $g^{RCP}_{(\pm1,\pm1)}\sim$ 300 meV . The obtained branches dispersions are shown in figure 3a as white dashed curves. Corresponding LCP data are reported in Figure S5d.

The c-SLR is a hybrid mode with variable plasmonic/photonic content depending on the coupling conditions tuned by the LP. This can be visualized through the Hopfield coefficients calculated for the various branches observed in the simulated extinction maps of RCP and LCP light (Figure S5).

To better evidence the photonic behaviour of the hybrid mode, in figure 3b we show that for detuning different from zero the Q factor increases, a trend that is also followed by the experimental values calculated in the nanofabricated arrays and reported in figure S6.

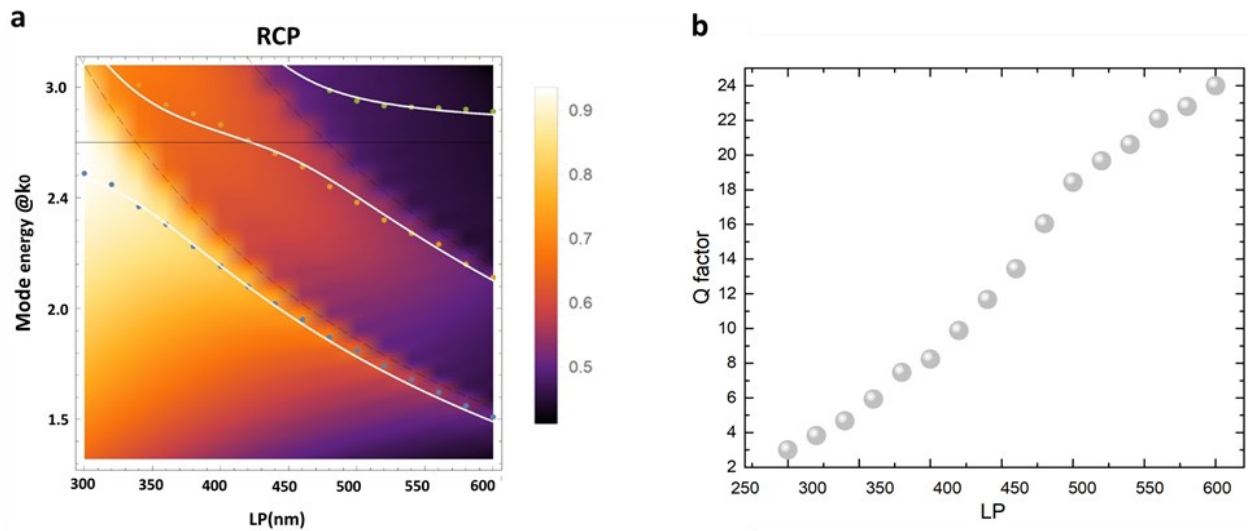

Figure 3. a) Extinction dispersion at normal incidence for right-handed circularly polarized light in water environment obtained from numerical simulations as a function of LP (from 300nm to 600nm). The solid black line indicates the calculated LSPR, the dashed black line indicates the calculated dispersions while the white solid lines plot the fitting of the calculated c-SLR peaks through the three coupled-oscillator model given by eq. 2, associated to the lower (blue points), middle (orange points), upper (green points) branches. b) Calculated Q factor ($\frac{\lambda}{\Delta\lambda}$) for the simulated first order c-SLR, as a function of the lattice period for RCP polarization.

As a benchmark of the proposed system towards sensing applications, we recorded the energy position of maximum CD in known refractive index environment (glycerol-water mixtures from 0% to 20% corresponding to a refractive index range between 1.333 and 1.358[31]) for the fabricated samples with different lattices. This spectral feature linearly shifts with the refractive index (figure 4a and S7) providing a sensitivity value, S, defined as Δλ/Δn (where Δλ represents the wavelength peak shift and Δn the change of the refractive index of the glycerol-water solution).

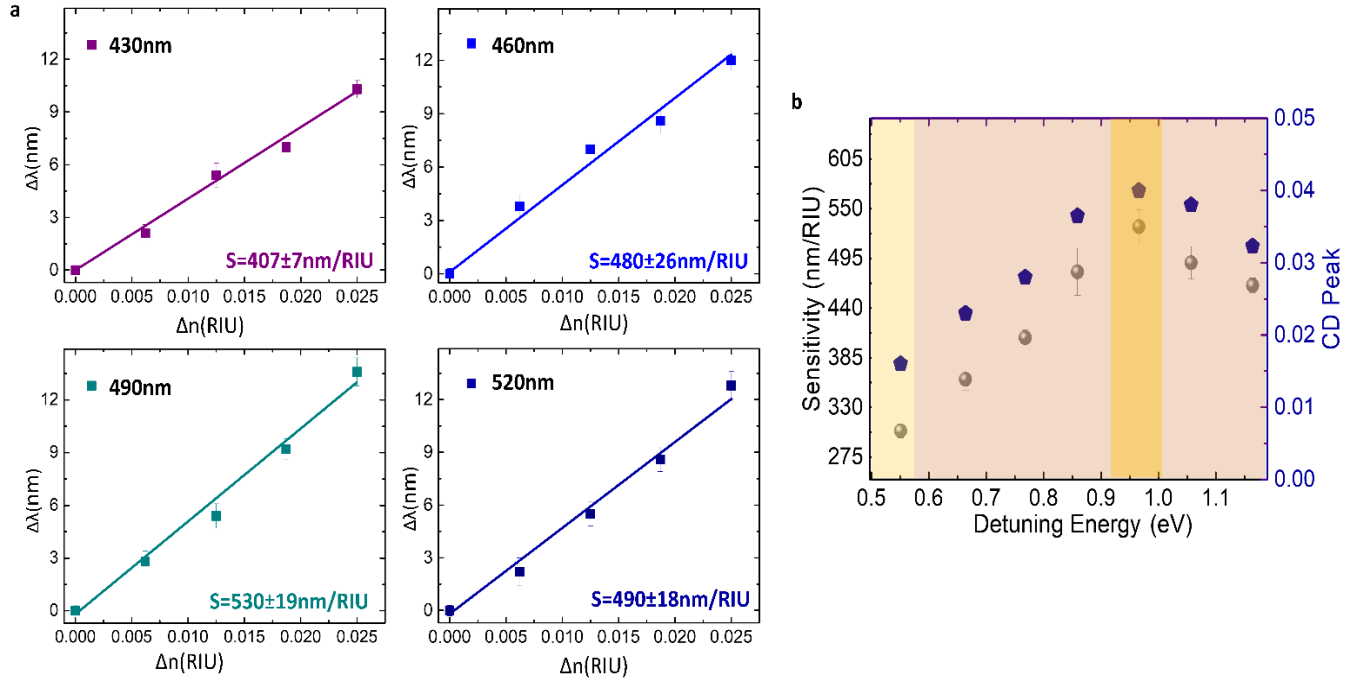

Figure 4. a) Trend of the spectral peak shift vs refractive index variations for nano-helices arrays with different LPs immersed in a glycerol-water solution at different molar concentrations. The line corresponds to the linear fits of the measured spectral positions. The sensitivity, measured as S=Δλ/Δn, is retrieved by the linear fit. b) Correlation between the sensitivity S and the maximum peak intensity of the CD (corresponding to the c-SLR peak). Both parameters increase up to a maximum value found at LP 490nm (corresponding to a 0,95 eV detuning between $E_{LSPR}$ and $E_g^{LP}$) and then decrease.

As shown in figure 4a, the sensitivity S evolves from 360 nm/RIU for small LP, up to the maximum value of 530nm/RIU for LP= 490nm and then decreases again to 490 nm/RIU for higher LP. Figure 4b highlights the direct correlation between S and CD experimental trends as a function of LP. Following the discussion above, the sensitivity trend could be explained considering the evolution of the plasmonic contribution to the hybrid mode. In the short LP region, the strong coupling between the plasmonic and photonic modes leads to the resonance splitting and to the suppression of radiative efficiency at such energy[32]. In this condition, characterized by a large plasmonic fraction, there is a large EM field confinement, leading to small modal volume V, but not enough to balance the low Q (figure 3b), thus resulting in less efficient sensing.

On the other hand, for larger detuning (large LP) the Q factor increases due to a more photonic behavior (Figure 3b)[32,33]. Therefore, to maximize the sensitivity, it is necessary to find an optimum balance between small chiral plasmonic mode volume (V) and large DOs Q factor (both influencing the field enhancement M proportional to $Q/\sqrt{V}$ [18]) in the c-SLR (figure S5). A sufficiently low plasmonic fraction can ensure the non-null and asymmetric scattering with respect to CPL, favoring a large sensitivity is achieved in combination with large CD value. Further increment of the LP>490nm inverts the trend, towards a reduction of the sensitivity, because despite of the increment of the Q factor (figure 3b), the further reduction of the plasmonic fraction corresponds to a low electric field enhancement (M). Following these results, we defined an intermediate LP region where the optimized Q/V ratio allows to reach high sensitivity S in the system.

In order to position our c-SLR sensor with current state of art in photonic and plasmonic metasensors[34], even though a wide plethora of architectures can be considered, we can make a comparison with the highest sensitivity system (to the best of our knowledge) based on periodic array of nanostructures exhibiting surface lattice resonance (S=357nm/RIU[35]) or on chiral nanostructures in solution with circular dichroism[36] (S=1,091nm/RIU, in the near-infrared). The obtained results support the large potential held by c-SLRs for high sensitivity purposes, especially considering the possible improvement to the technology related to material composition[37–39], helix architecture[40], or even the more challenging array size.

To conclude, we have shown the coupling between diffractive modes and chiral plasmonic resonances within a periodic array of fully 3D nanohelices. We have demonstrated that c-SLR can be excited in the visible spectral range on this system, provided that the criteria of sufficient size, homogeneous refractive index environment and lattice design are fulfilled. The coupling regime has been studied both theoretically and experimentally, as a function of the lattice period which regulates the features and the dispersions of CPL extinctions, as well as the relative percentage of photonic and plasmonic mode in the hybridized c-SLR. The sensing capability of the system has been also studied, with respect to surrounding medium refractive index tracking through the CD spectra. It is found that the sensitivity to refractive index changes in the environment can be maximized thanks to a trade-off between the photonic nature and the plasmonic losses in the hybrid mode (c-SLR). In particular, a small plasmonic fraction of the hybrid mode is desirable to preserve the chiro optical features of the system. Along with biosensing applications, the investigated mechanism can have several implications in miniaturized and integrated photonics, especially with respect to polarization control in optical emission.

**Experimental Section**

*Sample Fabrication.* The platinum-based nano-helices arrays have been grown on an ITO-on-glass substrate by means a Carl Zeiss Auriga40 Crossbeam FIB/SEM system coupled with a gas injection system (GIS) and trimethyl(methylcyclopentadienyl)platinum(iv) has been used as gaseous precursor.

The helix nanofabrication is obtained by Ga+ beam with energy at 30 keV, the beam current at 1pA, and the step size, at 10nm. The chamber pressure was kept between $8 \times 10^{-7}$ mbar and $1.06 \times 10^{-6}$ mbar during the process. 900 helices are replicated as in[22] according a predefined lattice design. During the growth, the proximity effects and the local pressure variation, that can affect the final size of each structures, has been controlled inserting a refresh time of 5 minutes each 30 elements, in order to keep the same values in the vacuum chamber and avoid the pressure drop.

*Optical characterization.* Extinction spectra both in real and the Fourier space imaging were recorded by using a homemade confocal setup made of an optical microscope Zeiss Axioscope A1 with a spectrometer. The sample was illuminated by circularly polarized light and focalized with a condenser with variable numerical aperture (from NA<0.1 to 0.95). The light was the collected using a 40x, NA=0.95 objective lens. Subsequently, the light passes through a three lenses system: the first reconstruct the real space, the second collimate the light beam and the third lens refocus the image in the real space. The light is then directed to a Hamamatsu Orca R2 CCD camera and a 200 mm spectrometer. By using all the three lenses combined with adjustable squared slits, the image can be selected in space. Removing the intermediate lenses, instead, Fourier space imaging has been obtained. The circularly polarized light has been produced using a linear Polarizer (Carl Zeiss, 400-800nm) and a superachromatic waveplate (Carl Zeiss, 400-800nm).

*Mode Fitting.* The maximum values of the normal incidence extinction and CD spectra were extracted by multiple Lorentzian fittings. The dispersions were fitted by using the solution for the eigenvalue problem of the Hamiltonian in eq.2 for both incident circular polarizations and CD.

*Numerical simulations.* Numerical simulations were performed by exploiting FTDT Based software. We have considered a single nanohelix and employed periodic boundary (Block) conditions in the array directions both to simulate extinction spectra and for energy−momentum extinction maps $E(k_x, k_y = 0)$ for different lattice periods. For the optical properties of the nanohelix, we considered a effective permittivity dispersion of Pt/C alloy calculated in our previous work[23]. The ITO/glass substrate was not included in the simulations. The surrounding environment changes around the nanohelices have been considered embedded the nanostructures array in a nondispersive dielectric media with different refractive indices.

**Acknowledgements**
M.M. and V.T. contributed equally to this work. The work was supported by "Tecnopolo per la medicina di precisione" (TecnoMed Puglia) – Regione Puglia: DGR no. 2117 del 21/11/2018 CUP: B84I18000540002 and Progetto PON ARS01_00906 "TITANNanotecnologie per l'immunoterapia dei tumori", finanziato dal FESR nell'ambito del PON "Ricerca e Innovazione" 20142020Azione II-OS 1.b).

**Supporting information**

**S1. Effect of the Array Size**

In order to give rise to the c-SLRs, another criterion is represented by the element number of the array[1,2]. In figure S1, we show how the observation of chiral SLR is reduced when considering an array of 20x20 elements (for a patterned area of 100 µm$^2$), rather than an array of 30x30 elements (for a patterned area of 225 µm$^2$). The nanofabrication of precise 3D structures is still a challenging nanofabrication issue. Our approach involves FIBID technology suitable to develop, in a bottom-up single direct writing step, more complex and fully 3D structures with high precision and reproducibility.

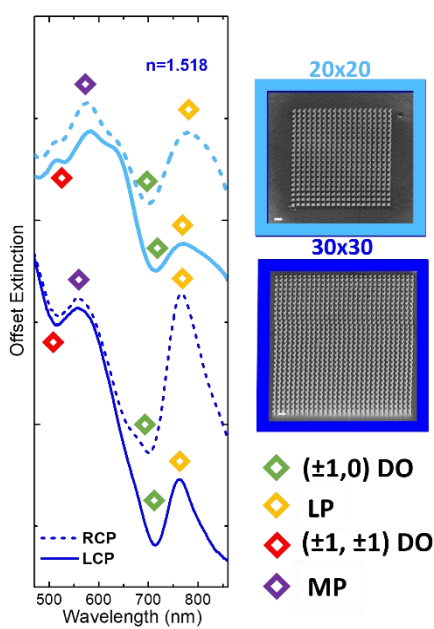

Figure S4. Comparison between the left and right-handed circularly polarized extinction spectra derived from transmission measurements of helix array (LP 460 nm) in oil (n=1.518) with different number of elements: 30x30 elements corresponding to a patterned area of 225 µm$^2$ (blue), and 20x20 elements corresponding to a patterned area of 100 µm$^2$ (light blue) showed in the left panel SEM image with scale bar 2 µm. The symbols with the same colors indicate similar spectral features (displayed in the legend) observed in both arrays.

**S2. Effect of Lattice Period**

Circularly polarized extinction spectra have been extracted from the central pixel of the measured extinction maps of the arrays of 30x30 elements with four different lattice periods (400nm, 430nm, 460nm, 490nm) displayed in the main text (figure 2). In periodic arrays of plasmonic chiral nanostructures, the chiral LSPRs can couple to the diffractive lattice modes of the array, generating the c-SLRs. This results in a peak in the extinction spectra (circle in figure S2a) preceded by a dip related to the linear dispersion of the (±1,0) diffractive orders. The extinction intensity increases with LP, and it is found to be higher for RCP, when incident polarization matches the same handedness of the helices, while is lower when interacting with the opposite light handedness. When increasing LP, a redshift of the spectral features and a progressive spectral narrowing is observed. For LP 460nm and LP 490nm, two additional modes are observed, corresponding to the coupling with (±1,±1) diffractive orders. This confirms that moving towards configurations different from the zero-detuning strong coupling regime, the CD increases [3].

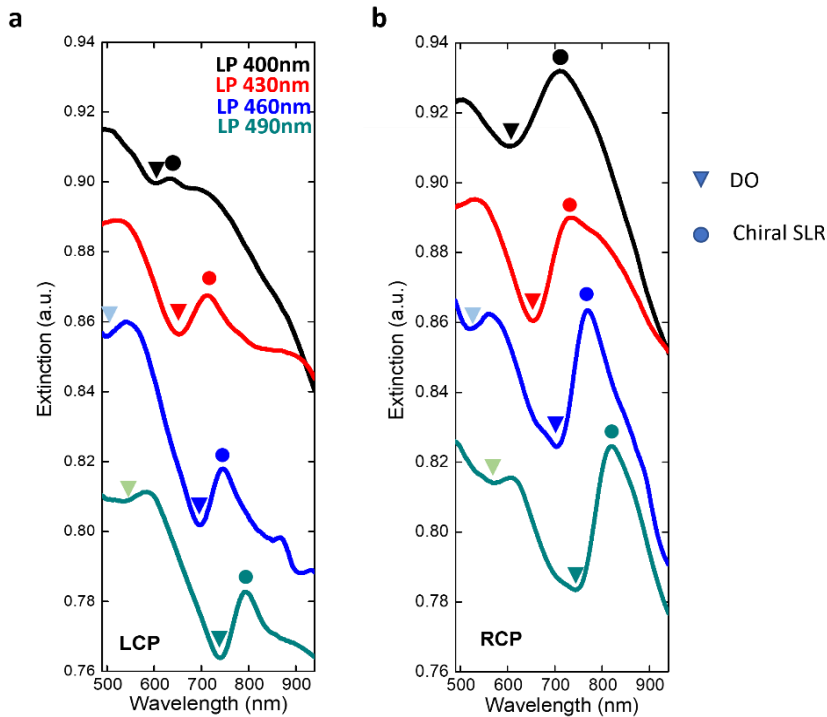

Figure S2. a-b. LCP and RCP extinction spectra recorded for arras of 30x30 elements and with different LPs in oil environment. The maxima are attributed to the c-SLRs spectral peaks (indicated by the circles), while the dips indicate the diffractive (±1,0) DOs (represented by the triangle). In the arrays with LP=460nm and LP=490nm, the lightest triangles indicate the (±1,±1) DOs.

**S3. Circularly Polarized Light Extinction Maps**

In figure S3 we display the measured LCP, RCP extinction maps and the retrieved CD. The c-SLR onset is evident for both the two incident polarizations, but with different intensity due to a more efficient dipole excitation in RH structure from RCP incident light. This is confirmed in the CD maps where chiral SLR signatures can also be observed. Their angular dispersion in the maps can be clearly distinguished confirming the hybrid nature of the c-SLR. In addition to what discussed in the main text, we highlight that the lower polariton branch (UP) mode superimposes with the (±1,±1) DO generating a new mode splitting (W(±1,±1) at 510 and MP at 560nm) for LP =490 nm.

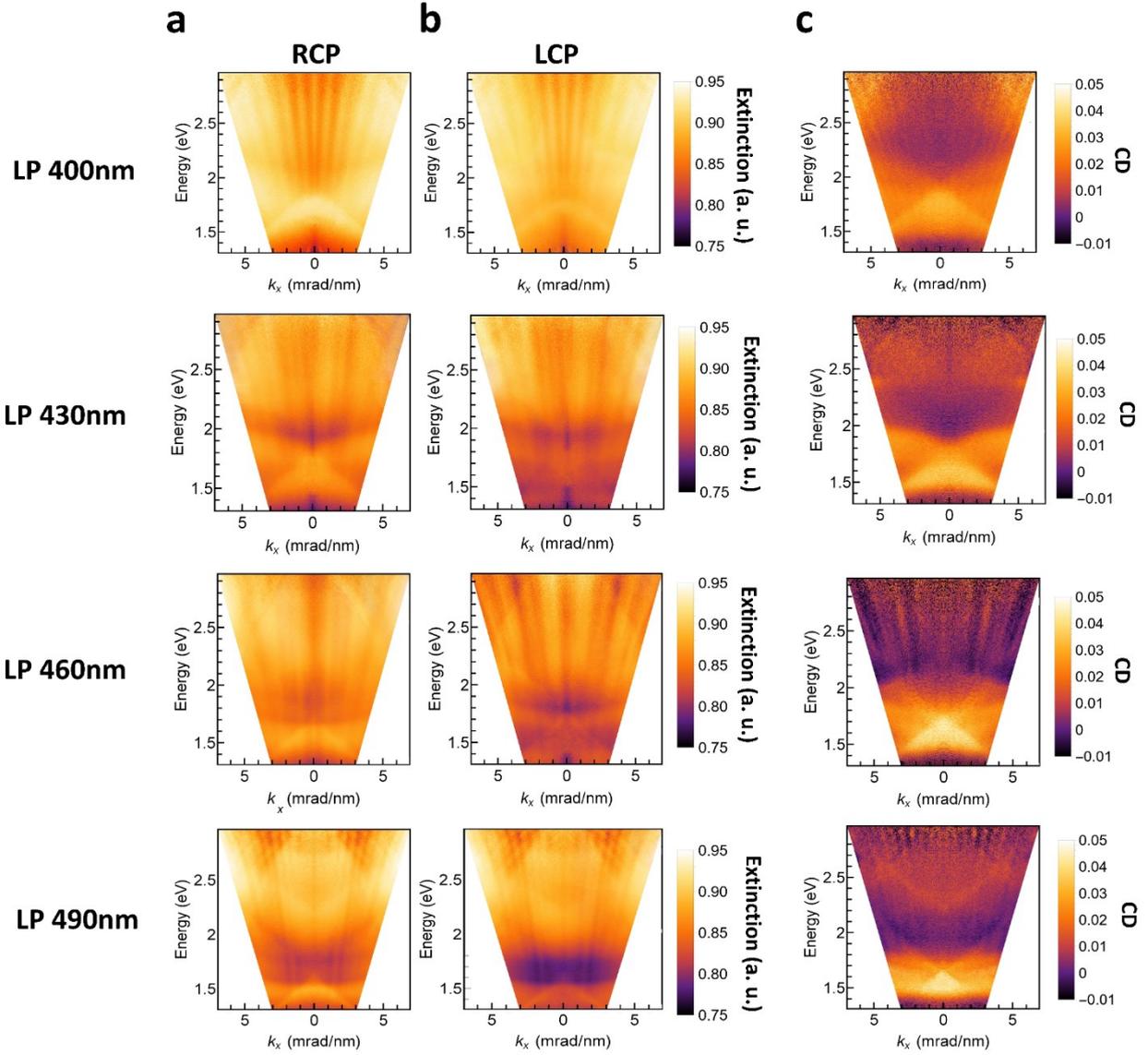

Figure S3. a-b. Experimentally measured energy-momentum extinction dispersions for LCP and RCP light for four different lattice periods (400 nm, 430 nm, 460 nm, 490 nm). c. Circular dichroism far-field maps as a function of LP calculated as the difference between RCP and LCP extinctions.

## S4. Numerical simulations of the energy-momentum far field extinction dispersions

The simulated extinction far-field maps of figure S4 are retrieved under interaction with LCP and RCP light in oil environment for different lattice periods: 400nm, 430nm, 460nm, 490nm. They present a good agreement with what observed experimentally and reported in the main text. In particular, we observe a similar trend of the c-SLR with the LP variation with the experimental data. RCP extinction maps display higher intensity, suggesting a better spectral overlap when matching the polarization and the structural handedness, according to experimental data.

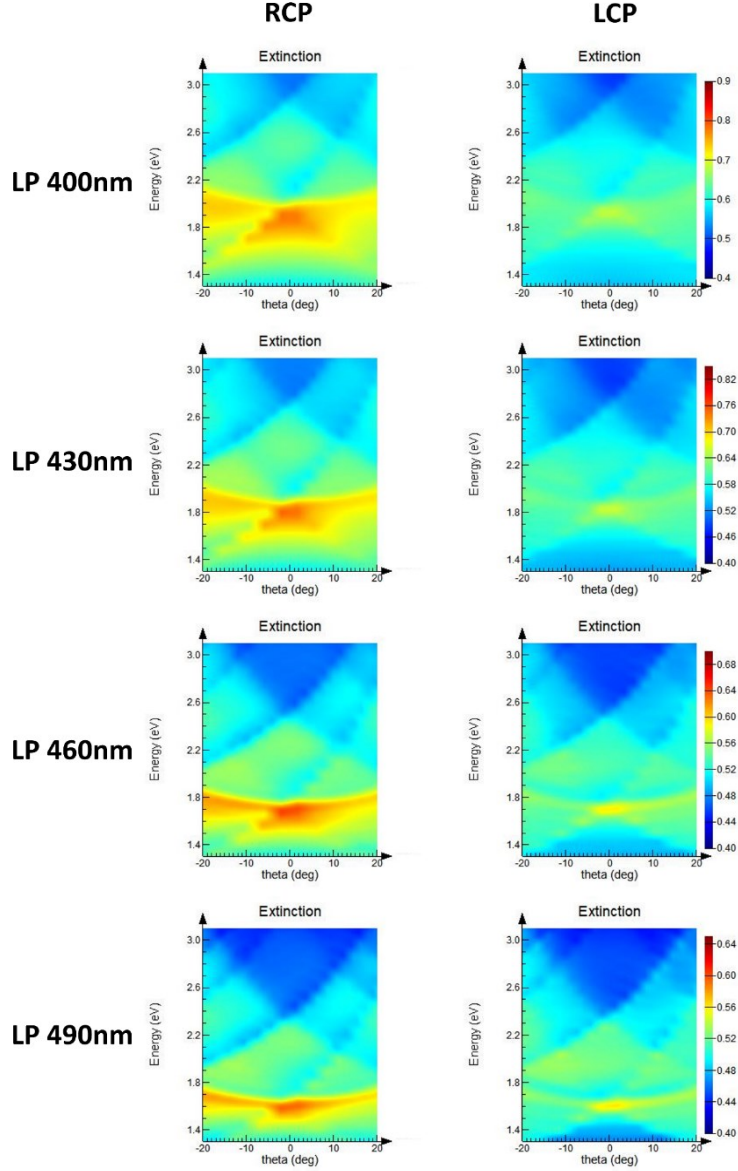

Figure S4. RCP and LCP simulated far-field extinction maps evaluated in oil environment for different LPs.

## S5. Calculated Hopfield Coefficients

Figure S5 a,b,c display the Hopfield coefficients for the upper, lower and middle branches, respectively, referred to RCP incident light condition discussed in the main text (figure 3 a). The plots describe how the photonic/plasmonic fraction of the hybrid mode changes with LP.
The Hopfield coefficients ($\alpha$) can be calculated as in [4]:

$$\alpha = \frac{1}{\sqrt{2}} \frac{\frac{\Delta}{2} + \Omega}{\sqrt{\left(\frac{\Delta}{2}\right)^2 + g^2 + \frac{\Omega\Delta}{2}}}$$

Where $\Delta$ is the detuning at a given lattice k-vector, $\Omega = \sqrt{\left(\frac{\Delta}{2}\right)^2 + g^2}$ and g is the coupling coefficient, that is half the separation between the lower and upper branches at zero detuning.
In the lower branch, for low LPs (<340nm) the hybrid mode is mostly plasmon-like. Increasing LP, the plasmon mode fraction decreases and the DO(±1,0) mode weight becomes predominant. At LP = 340 nm, that is the zero detuning condition, the plasmonic and the DO(±1,0) fractions are the same.

Analogously, in the upper branch(figure S5b) for low LPs (<440nm) the mode is almost pure DO-like, while increasing LP, the DO mode fraction decreases and the plasmon fraction increases; for higher LP, the latter value becomes predominant. The middle branch composition (Figure S5c) is more complex since it is weighted over the plasmonic and the different DOs contributions.

Under LCP illumination, the calculated extinction dispersion (figure S5d) shows a plasmon energy extracted at 2.9eV (indicated by a solid line), while the uncoupled dispersions are represented by the black dashed lines. The solid white lines plot the fitting of the simulated peaks through the three coupled-oscillator model given by equation 2. The coupling strength gave values of $g_{[\pm1,0]} \sim 270meV$ and $g_{[\pm1, \pm1]} \sim 170meV$. For the Hopefield coefficients, in all the calculated branches a behaviour similar to RCP excitation is observed (Figure S5e).

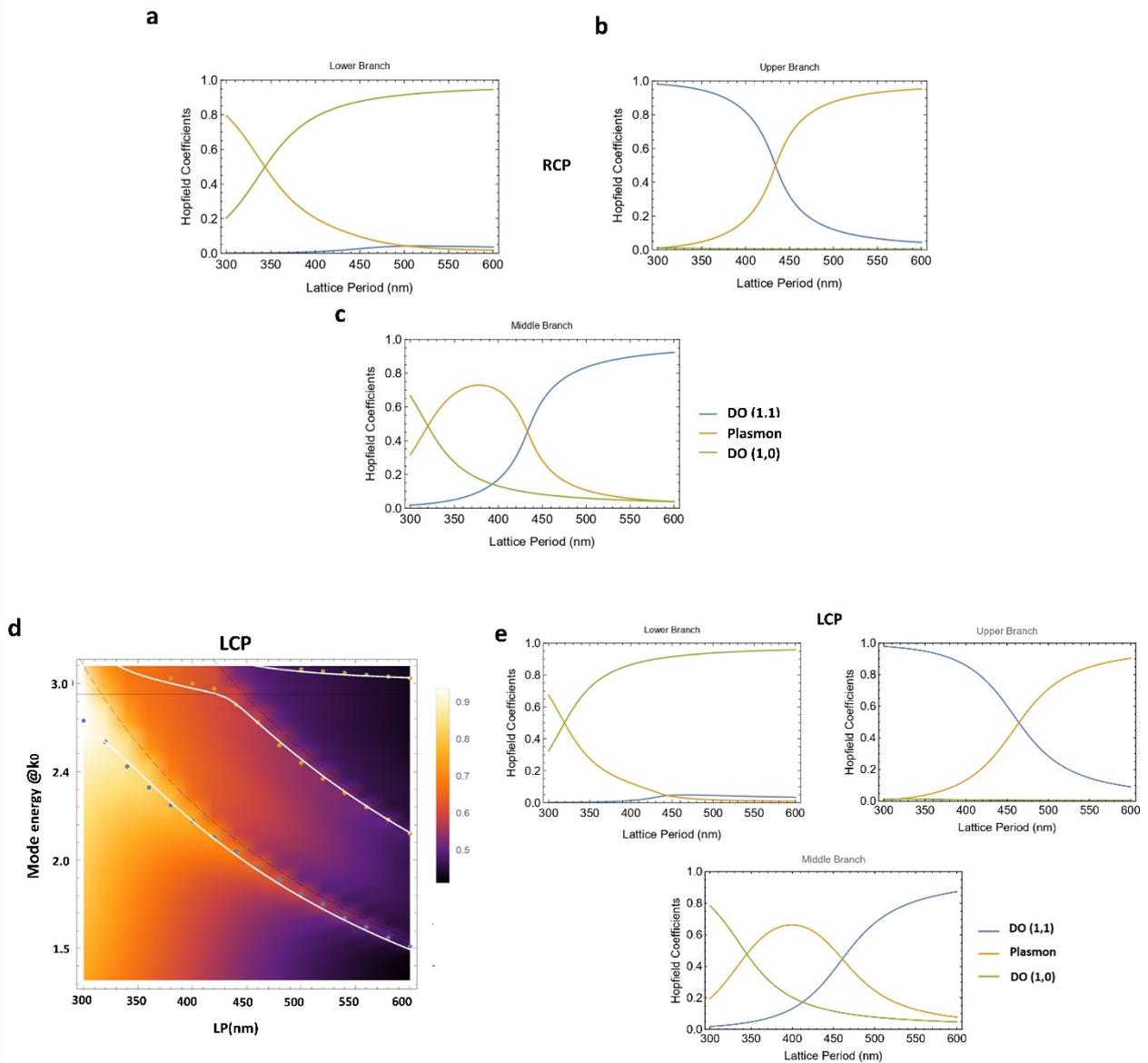

Figure S5. Hopefield coefficients calculated from RCP extinction maps for the lower (a), upper (b), and middle (c) branches. d) Extinction dispersion at normal incidence for left-handed circularly polarized light in water environment obtained from numerical simulations. e) Hopefield coefficients for LCP retrieved from the simulated extinction dispersion.

**S6. Q-factor**

The Q-factor of figure S6 was calculated from the experimental spectra acquired in water environment (n=1.3334) as:

$Q = \lambda/\Delta\lambda$,

where λ corresponds to the spectral peak of c-SLRs, and Δλ is the full width at half maximum derived by the Gaussian fit of the SLR spectra. The measured Q-factor shows a clear increase with the lattice period, caused by a reduction of plasmonic content. It should be noted that the material composition of NH fabricated by FIBID is not purely metallic, thus explaining the low absolute values of measured c-SLR Q-factor.

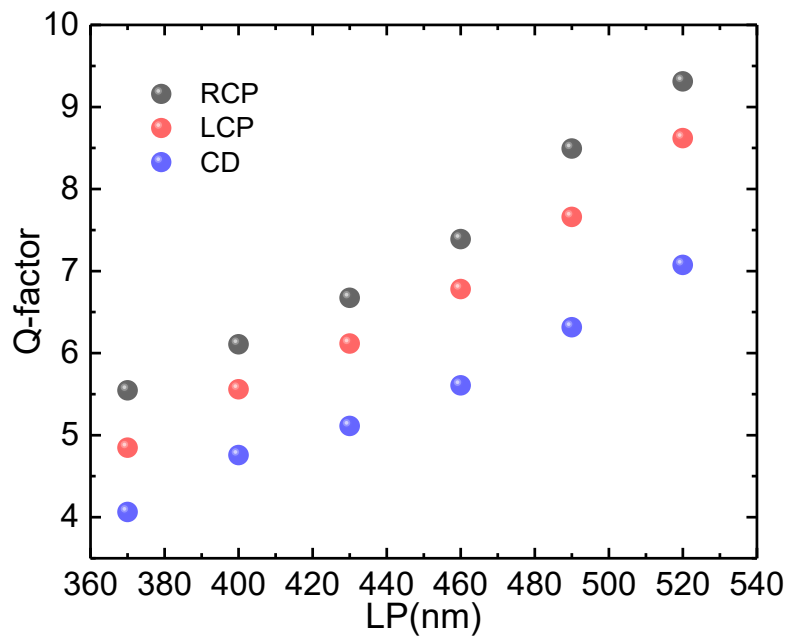

Figure S6. Q- factor values calculated for LCP, RCP and CD extinction measured in water environment (n=1.3334) for different LPs.

**S7. CD sensitivity to refractive index variations**

The CD spectra of the samples with different lattice periods were collected in known refractive index (RI) environment. We considered glycerol–water mixtures with varying concentration from 0 to 20% (corresponding to a refractive index range between 1.333 and 1.358)[5]. All the spectra red-shift and, in particular, we focused on the shift of CD maxima, shown in the panels of figure S7, which were used to calculate the sensitivity values in nm/RIU shown in the main text, figure 4a.

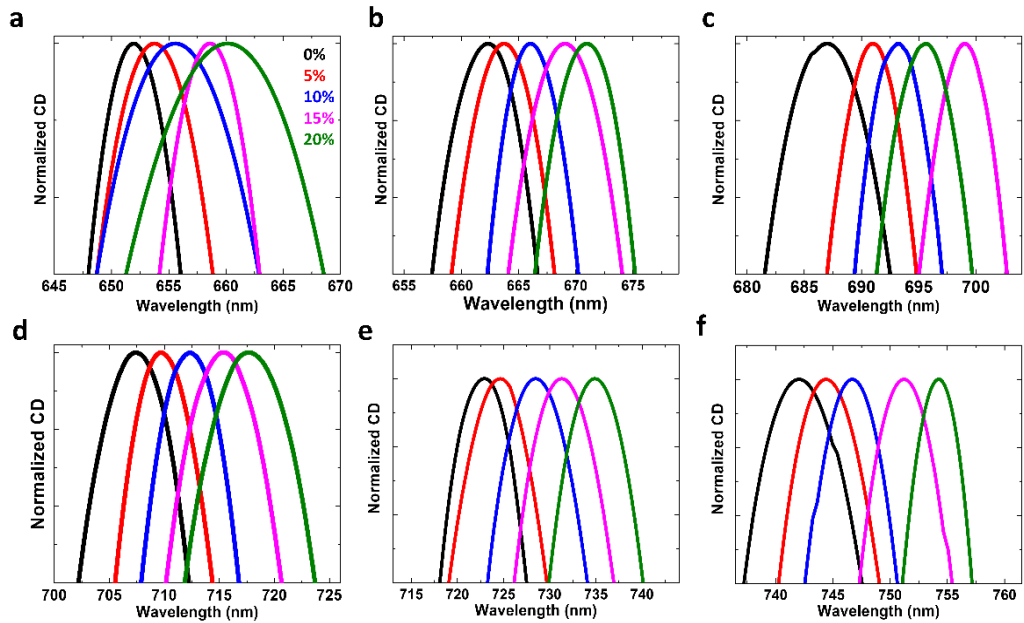

Figure S7 a-f. CD spectra of the NH arrays as a function of the surrounding refractive index environment (0%, 5%, 10%, 15% 20% of glycerol in water solution), zoomed at the relative resonance peaks measured for different lattice periods: a)LP 370nm, b)LP 400nm, c) LP 430nm, d) LP 460nm, e) c) LP 490nm, f) LP 520nm.